# Ultra-broadband integrated optical parametric amplifier for quantum sensing

Kai-Chi Chang[1,*], Tushar Sanjay Karnik[1,3,*], Chun-Ho Lee[1,2], Kiyoung Ko[1], Xinyi Ren[1,2], Ian Christen[1], Reshma Kopparapu[1,2], Clayton Cheung[1,2], Kiwon Kwon[1], Kamila Kunes[1,2], Zaijun Chen[1,2], and Mengjie Yu[1,2,3,†]

[1]Department of Electrical Engineering and Computer Sciences, University of California, Berkeley, CA 94720, USA

[2]Ming Hsieh Department of Electrical and Computer Engineering, University of Southern California, Los Angeles, CA 90089, USA

[3]Materials Sciences Division, Lawrence Berkeley National Laboratory, Berkeley, CA 94720, USA

*These authors contributed equally.

†Corresponding author: mengjie.yu@berkeley.edu

**Abstract**

Although thin-film lithium niobate (TFLN) facilitates efficient signal generation and nonlinear and quantum interactions, the realization of optical parametric amplification (OPA) that can provide simultaneous ultra-broadband and high-gain operation in an integrated chip continues to pose a challenge. Here we demonstrate continuous-wave-pumped OPA in an X-cut MgO-doped, dispersion-engineered, and adaptively-poled TFLN waveguide of 1.6 cm length, achieving a flat-top profile covering a 450 nm-wide optical wavelength window, corresponding to a 3-dB gain bandwidth of about 56 THz. Among reported TFLN platform, our device exhibits the broadest 3-dB gain bandwidth. The same device can be pumped either directly at visible wavelengths through second-order $\chi^{(2)}$ interactions or in the telecom band through cascaded $\chi^{(2)}$ processes, the latter eliminating the need for a high-power visible pump laser. We achieve maximum gains of 8.87 ± 0.39 dB and 10.79 ± 0.43 dB at on-chip pump powers of 74 mW and 170 mW for the direct and cascaded schemes, respectively. We further directly probe OPA gain across the 1650-1900 nm wavelength range, where experimental gain measurements have remained scarce. With a normalized on-chip gain of 1.21 dB/(W·mm), our device sets a new benchmark among reported cascaded $\chi^{(2)}$ nonlinear processes. This work advances the realization of integrated optical parametric amplifiers, offering high efficiency, robust gain, ultra-broadband bandwidth, and continuous-wave operation, thereby enabling new capabilities for next-generation quantum sensing and photonic systems.

## 1. Introduction

Optical amplification underpins much of modern information infrastructure, making possible long-distance optical transmission, worldwide fiber-optic networks, and a range of present-day applications spanning satellite links to data-center interconnects. Even though this technology is well established, there is increasing demand for amplification beyond the standard erbium band, motivated by S-, C-, and L-band multiband transmission for high-capacity telecom systems, O-band links for data centers, long-wave infrared applications [1], and near-infrared/visible-wavelength quantum applications.

Practical deployment of fiber-based OPAs, however, remains difficult, since they call for hundreds of meters of highly nonlinear fiber and require suppression of Brillouin scattering arising from the high pump powers involved [2]. Over the past few years, TFLN has become a prominent material platform for realizing OPA and squeezed-light generation, owing to its tight optical mode confinement and strong second-order nonlinear response [3, 4]. For example, squeezed light has driven major sensitivity gains in gravitational-wave detectors over the past decade [5], underscoring its value for quantum sensing [6-16]. It has also advanced performance in dual-comb spectroscopy [17], optomechanical sensing [18], and quantum computing [19-22]. By confining light to a much smaller mode field size, TFLN devices achieve substantially stronger nonlinearity than bulk materials [23-27], opening up new possibilities for high-gain OPAs that operate with significantly reduced peak power, for instance, using continuous-wave (CW) pumping. Traditionally, achieving parametric gain via $\chi^{(2)}$ nonlinear processes requires direct pumping at twice the fundamental frequency or a separate nonlinear crystal to generate a strong second-harmonic generation (SHG). A more elegant approach relies on cascaded $\chi^{(2)}$ nonlinear processes. This method enables the use of well-developed telecom-band pump sources, such as erbium-doped amplifiers, rather than requiring direct pumping with visible lasers. OPAs based on both Kerr [28-30] and second-order nonlinearities [31] have recently been shown in various photonic integrated platforms, including TFLN, silicon nitride, gallium phosphide, and thin-film lithium tantalate, showing clear advantages over their fiber-based counterparts [32].

In this work, we present a CW-pumped OPA implemented on the TFLN platform. We employ both visible direct pumping and cascaded SHG and OPA processes with indirect near-infrared (NIR) pumping. We measure on-chip CW flat-top parametric gain across an ultra-broadband bandwidth of 450 nm, equivalent to about 56 THz, which is more than an order of magnitude wider than the bandwidth of conventional erbium amplifiers and broadest bandwidth demonstrated on the TFLN platform. This is accomplished through precise higher order dispersion engineering of the integrated TFLN device's gain bandwidth, extending it beyond the S-, C-, and L-band telecommunication wavelengths [33]. Under visible pumping, the maximum gain reaches 8.87 ± 0.39 dB at an on-chip pump power of 74 mW. In contrast, NIR pumping yields a maximum gain of up to 10.79 ± 0.43 dB at an on-chip pump power of 170 mW. This is enabled by harnessing the second-order nonlinearity of quasi-phase-matched (QPM) [34, 35] waveguides built on the recently developed adaptively poled TFLN platform [15]. Additionally, we characterize the OPA gain over the 1650-1900 nm range, a region where direct gain measurements have so far been missing. This work achieves a normalized on-chip gain of 1.21 dB/(W·mm) for cascaded $\chi^{(2)}$ nonlinear processes, the highest reported to date [36-41]. These findings mark a meaningful advance toward practical, photonic integrated OPAs that operate in a low-noise CW regime, handle sufficient input powers, and deliver ultra-broadband flat-top gain for quantum sensing and photonic systems.

## 2. Results

As depicted in Figure 1(a), we present the schematic of OPA processes in a single-pass TFLN device, with visible and NIR pumps, probe light evolution along the waveguide. A two-photon microscopy image of the PPLN waveguide, with uniform domain inversion across its entire length. Our method uses TFLN waveguides, 16 mm in length, fabricated with a 602-nm-thick, MgO-doped, X-cut TFLN wafer [15]. We use adapted poling techniques [15, 25] to address the issue of limiting nonlinear conversion efficiency. The PPLN waveguide width is set to 1.9 µm to minimize propagation loss across both telecom-band and visible wavelengths. Figure 1(b) and 1(c) are the energy diagrams illustrating $\chi^2$ and cascaded $\chi^2$ processes. Then, in Figure 1(d) and 1(e), for visible direct pumping and NIR pumping, the loss of the waveguide is measured to be 0.1 dB/cm and 0.4 dB/cm, respectively [15, 16]. In Figure 1(d), purple lines are the visible direct pumping with 170 mW power, while Red lines are the visible direct pumping with 74 mW power. For visible direct pumping, the on-chip power decreases with the propagation loss of the waveguide, and the higher on-chip power will provide higher parametric gain as compared to the NIR pump case. In Figure 1(e), for NIR pumping, this cascaded approach allows the use of SHG generation and then OPA process in a same waveguide. Blue line is the power for NIR pumping, and this approach allows to use mature telecom-band pump sources and amplifiers instead of direct visible pumping. With this scheme, the input coupling only needs to be optimized for the telecom-band wavelengths.

Green line is the building up SHG visible power in cascaded $\chi^2$ processes. For visible direct pumping and NIR pumping, the parametric gain amplification builds up while propagating through TFLN waveguides. Overall, visible and telecom-band pumping provide complementary advantages. Visible pumping is favorable for high-gain operation when sufficient pump power is available, facilitates phase-sensitive gain at degenerate fundamental frequencies, and enables straightforward spectral separation and filtering of the pump from the signal. In contrast, telecom-band pumping benefits from the availability of mature high-power sources and optical components, offers more robust power handling at the chip facets with reduced photorefractive effects, and eliminates the need for a high-power visible laser. In this work, we utilize both schemes to investigate the gain bandwidth and performance of OPA in our TFLN devices. We superimpose experimental maximum gain in Figure 1(d) and 1(e) with stars.

Figure 2(a) is the image of the OPA chip with the waveguide glowing under high-power NIR pumping. Enabled by the adaptive poling design, we tailor the local poling period to sustain strong parametric gain over the full device length, achieving an SHG efficiency of 3953.15%/W, as presented in Figure 2(b). In Figure 2(b) inset, SHG is measured with the telecom pump laser (Santec TSL 570). Lensed fibers are used throughout this work for coupling to the TFLN waveguides, giving average fiber-to-chip insertion losses of 4.2 dB (C-band) and 5.31 dB (780 nm). Figure 2(c) is the cross-section view of the TFLN waveguide. The device parameters are as follows: a waveguide width of 1.9 μm, a length of 1.6 cm, a side-wall angle of 60°, an etch depth of 356 nm, an oxide thickness of 100 nm, and a total device thickness of 602 nm. In Figure 2(d), we present the simulated dispersion coefficients $\beta_2$, and $\beta_4$ for our TFLN waveguide. By fitting the dispersion curves, we estimate the dispersion coefficients as $\beta_2 \approx 3.357$ fs$^2$/mm, and $\beta_4 \approx 1143.032$ fs$^4$/mm, at fixed pump wavelength as the green line. Figure 2(e) shows the simulated 3-dB gain bandwidth for our TFLN device, yielding a broad, flat-top gain spectrum. The simulated OPA 3-dB gain bandwidth is 454 nm. These simulations were carried out using device parameters in Figure 2(c). Inset of Figure 2(e) is the total phase-mismatch simulation. Since the 3-dB gain bandwidth of the OPA is governed by the phase mismatch in the three-wave-mixing process, we plot the total phase mismatch using the same device parameters. The combination of near-zero second-order dispersion $\beta_2$ and fourth-order dispersion $\beta_4$ is what enables the flat, broadband gain profile at a fixed pump wavelength. The dispersion relation can be summarized in the equation as follows: $\Delta\beta \approx \beta_2(\omega_p - \omega_S)^2 + \frac{1}{12}\beta_4(\omega_p - \omega_S)^4$, where $\omega_p$, and $\omega_S$ are the pump and signal angular frequencies, respectively. Then, we proceed to OPA gain bandwidth measurement with NIR pumping. We use an L-band amplifier cascaded with a tunable bandpass filter (OZ Optics) to amplify the telecom-band pump while suppressing the background amplified spontaneous emission noise from the L-band amplifier. The output from TFLN waveguide is routed to an OSA (Yokogawa AQ6370D), set to 0.1 nm resolution bandwidth, chosen based on the wavelength range of interest. Figure 2(f) is the measured spontaneous emission spectrum of the integrated OPA, showing a 3-dB bandwidth of 450 nm for the TFLN amplifier, which is consistent with our simulation results in Figure 2(d). In addition, a comparison of 3-dB gain bandwidths for recent integrated OPAs versus prior literature is provided in Supplementary Material Figure S1. Our 3-dB gain bandwidth is the broadest for TFLN platforms [33, 36].

With this, we proceed to the OPA gain experiments via visible direct pumping at 784.45 nm, schematically shown in Figure 3(a). The OPA chip sits on a temperature-controlled metal mount with a stability of 2 mK. We use a Booster TA to amplify the visible pump laser, and use off-chip WDMs to combine and separate the strong visible pump from the probe light before and after the chip. The probe light is set to -30 dBm for parametric gain measurements. In our measurements, the probe light consists of O-band (Santec 570), S-, C-, L-band (Santec 570), and in the 1650-1900 nm (Radiantis) range, respectively. The 780 nm port from output WDM is used to monitor the visible pump light, while the other port is routed to the OSA for the gain measurements. For the following gain measurements, we keep the 24 dBm of off-chip visible power, with the on-chip visible power reaching 74 mW. This power is chosen to have better stability for gain measurements. Figure 3(b) to 3(d) shows the measured raw optical spectra of the TFLN waveguide chip with the visible pump on and off, at 1360, 1550, and 1690 nm, yielding measured gains of 1.45 dB, 8.52 dB, and 8.26 dB, respectively. Additional OPA gain measurements at other wavelengths are presented in Supplementary Material Figure S2. We record gain spectra at a single frequency across a 5 nm window centered on the probe light wavelength, using a resolution bandwidth of 0.1 nm. At 1500 nm, the smallest noise figure is 3.34 dB. Figure 3(e) to 3(f) are the gain measurements across the O-, S-, C-, and L-bands, and in the 1650-1900 nm range, the latter sampled more densely since OPA gain in this range has remained largely unreported [33, 36-41]. The experimental gain is calculated with $G = P_{on}/P_{off}$, where $P_{on}$ and $P_{off}$ are the measured output probe light powers with the visible pump on and off, respectively. From the SHG measurements in Figure 2(b), the theoretical maximum gain is 9.25 dB by solving coupled mode equations. Hence, our measurements agree well with theory and with the emission spectrum in Figure 2(f), reaching a maximum gain of 8.87 ± 0.39 dB. The theoretical calculation of maximum gain is shown in Supplementary Material Figure S3. Deviations at wavelengths beyond the 450 nm 3-dB amplification bandwidth account for the lower-than-predicted gain there. Error bars reflect the standard deviation across three measurements.

Next, we proceed to broadband amplification measurements in TFLN waveguides using the NIR pumping at 1568.9 nm. For this measurement, the WDMs are replaced with 90:10 fiber couplers to cover the widest possible bandwidth for the probe lights. Figure 4(a) is the experimental setup, where we use the L-band amplifier, tunable filter and OSA. In this setup, we keep the probe light at -20 dBm, to be consistent with the probe light power level in Figure 3(a). For subsequent gain measurements, we maintain the off-chip telecom power at 26.5 dBm, corresponding to an on-chip power of 170 mW, a level chosen for improved measurement stability. In Supplementary Material Figure S4, we provide SHG conversion efficiency measurements versus simulations under NIR pumping. The experimentally measured SHG conversion efficiency peaks at 87.59%, approaching the 88.43% theoretical limit under a similar level of NIR pump power input. This theoretical limit is derived from coupled mode equation simulations that account for waveguide propagation loss. Figure 4(b) to 4(d) are the measured raw optical spectrum of the TFLN waveguide chip with the telecom pump on and off, at 1360, 1550, and 1690 nm, with measured gains of 2.05 dB, 10.17 dB and 10.81 dB, respectively. Additional OPA gain measurements at other wavelengths appear in Supplementary Material Figure S5. To be consistent with our measurements in Figure 3, here the gain spectra are captured at a fixed frequency over a 5 nm span centered on the probe wavelength, with the resolution bandwidth set to 0.1 nm. At 1500 nm, the noise figure reaches a minimum of 3.48 dB. The higher pump power used here results in a slightly higher noise figure relative to visible direct pumping. Figure 4(e) to 4(f) are the gain measurements covering the O-, S-, C-, and L-bands, and the 1650-1900 nm range, the latter sampled more densely as before. We obtain the measured gain using $G = P_{on}/P_{off}$, where $P_{on}$ and $P_{off}$ are the measured output probe light powers with the telecom pump on and off, respectively. From the SHG measurements in Figure 2(b), the predicted gain is 11.21 dB. Our results agree with theory and with the emission spectrum in Figure 2(f), reaching a maximum gain of 10.79 ± 0.43 dB; the theoretical calculation of maximum gain is shown in Supplementary Material Figure S6. Lower-than-predicted gain at certain wavelengths reflects operation outside the 450 nm 3-dB amplification bandwidth. Error bars represent the standard deviation across three measurements. Here we also explain the background amplification observed in our gain measurements (Figures. 3(b) to 3(d), 4(b) to 4(f)). When the pump is on, the same nonlinear process that amplifies the probe lights also parametrically amplifies vacuum fluctuations at other wavelengths across the OPA's gain bandwidth, consistent with prior reports [33, 36, 41]. In Supplementary Material Figure S1, we compare the on-chip gain normalized by device length and on-chip pump power for our works and other recent works. This work features the highest on-chip normalized on-chip gain of 1.21 dB/(W·mm) for cascaded $\chi^{(2)}$ nonlinear processes.

**3. Conclusion**

In summary, via using TFLN waveguides, we demonstrate cascaded and direct second-order CW pumping OPA with a flat-top bandwidth as large as 450 nm. Our waveguides, just 1.6 cm long, are designed to operate in the anomalous but near zero group velocity dispersion regime, yielding continuously broad bandwidth without requiring phase-matching tuning via temperature or pump wavelength adjustment. Unlike narrowband OPAs in optical fibers and bulk crystals [32], which cannot be freely positioned across the spectrum and must operate near bands with available high-power pump sources, our TFLN OPA spans all standard communication bands and well beyond, removing the need for high-power sources outside these mature wavelength regions. Within this ultra-broadband 450 nm 3-dB gain bandwidth, and with an adapted poling technique to mitigate nanoscale inhomogeneities, we achieve maximum gains of 8.87 ± 0.39 dB and 10.79 ± 0.43 dB with low on-chip pump powers of 74 mW and 170 mW, respectively. Compared with previously reported TFLN platforms, our device demonstrates the broadest 3-dB gain bandwidth [33, 36]. Our work features record normalized on-chip gain of 1.21 dB/(W·mm) for cascaded $\chi^{(2)}$ nonlinear processes [36-41]. Notably, this performance is achieved with the simplest possible device architecture, single-pass straight waveguides. By forgoing complex photonic structures and cavity-based enhancement, the integrated OPA gains exceptional robustness and reliability for practical deployment. Furthermore, the OPA gain bandwidth in TFLN can further be dispersion engineered to cover larger wavelength ranges, making it a promising platform for broadband and two-mode squeezing protocols. For example, negative second-order dispersion coefficients $\beta_2$ can be achieved through slightly increasing the etch depth or decreasing the waveguide width, resulting in a broader gain bandwidth. Moreover, by lowering waveguide propagation loss for NIR and visible wavelength to around 1.3 dB/m [15], and 6 dB/m [42], it is possible to push the SHG efficiency toward its theoretical limit of 7774.72%/W. This could enable higher optical parametric gain of 19.73 dB under 170 mW NIR on-chip pump power. Additionally, phase-sensitive amplification [43-46], and on-chip squeezed light [6-16], could be further demonstrated to fully realize the high-SNR advantage below the standard quantum limit. The highly efficient, low-loss nonlinear integrated device demonstrated here serves not only as an essential OPA for photonics applications in optical communication [1], photonic processors and photonic computing [47], but also as a versatile tool for fundamental research spanning quantum optics [48], and precision spectroscopy [17].

**Data availability**

The datasets generated and analyzed during this study are available from the corresponding author upon reasonable request. Source data are provided with this paper.

**Acknowledgements**

This work is supported by the DARPA INSPIRED program (HR001123S0052), the DARPA Young Faculty Award (D23AP00252-02) and the Air Force Office of Scientific Research under award number FA9550-24-1-0349. Device fabrication was performed at the John O'Brien Nanofabrication Laboratory at University of Southern California, Marvell Nanofabrication Laboratory at University of California, Berkeley, University of California Los Angeles Nanolab and Molecular Foundry at Lawrence Berkeley National Laboratory. M.Y., and T.S.K. are supported by the U.S. Department of Energy, Office of Science, Basic Energy Sciences, Materials Sciences and Engineering Division under Contract No. DE-AC02-05CH11231 within the Quantum Coherent Systems Program KCAS26. The views, opinions and/or findings expressed are those of the authors and should not be interpreted as representing the official views or policies of the Department of Defense or the U.S. Government.

**Author contributions**

M.Y. conceived the idea. K.-C.C., and T.S.K., carried out experiments. K.-C.C., T.S.K., K.K., and X.R. analysis data and perform simulations. T.S.K. designed the chip with the help of R.K., C.C., C.-H.L.. C.-H.L. and T.S.K. fabricated the devices and developed the fabrication processes with the help of K.K.. I.C., and M.Y. supported and discussed the studies. K.-C.C., and M.Y. wrote the manuscript with contributions from all authors. M.Y. and Z.C. supervised the project.

**Competing interests**

C.-H.L., Z.C., and M.Y. are involved in developing lithium niobate technologies at Opticore Inc.

**Supplementary material** The online version contains Supplementary Material.

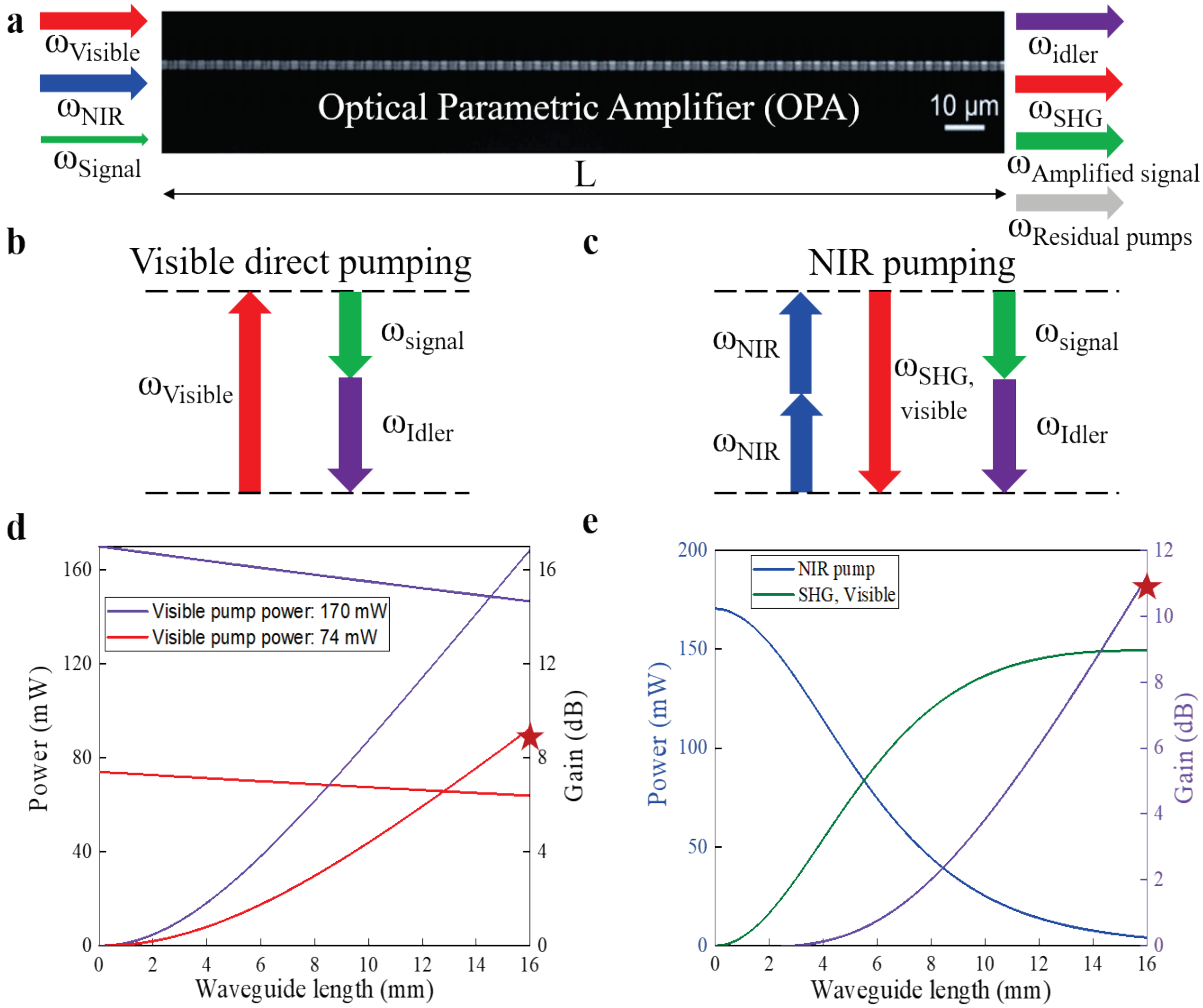

**Figure 1. Chip-scale OPA in thin-film lithium niobate (TFLN) waveguides. (a),** Schematic of OPA processes in a single-pass TFLN device, with visible and NIR pumps, probe light evolution along the waveguide. The device's periodically poled domains are shown with two-photon microscope image. **(b)** and **(c)**, Schemes for $\chi^2$ and cascaded $\chi^2$ processes. **(d)** and **(e)**, For visible direct pumping and NIR pumping, the loss of the waveguide is measured to be 0.1 dB/cm and 0.4 dB/cm, respectively [15]. **(d)**, Purple: visible direct pumping with 170 mW power, the same pump power used in the NIR pump case. Red: visible direct pumping with 74 mW power (used in our experiments). **(e)**, Blue: for NIR pumping, this approach allows to use mature telecom pump sources and amplifiers instead of direct visible pumping. Green: SHG power in cascaded $\chi^2$ processes. Purple: the gain from the cascaded $\chi^2$ processes. We can see the parametric gain amplification builds up while propagating through TFLN waveguides for both schemes. We superimpose experimental maximum Gain in panel **(d)** and **(e)** with stars.

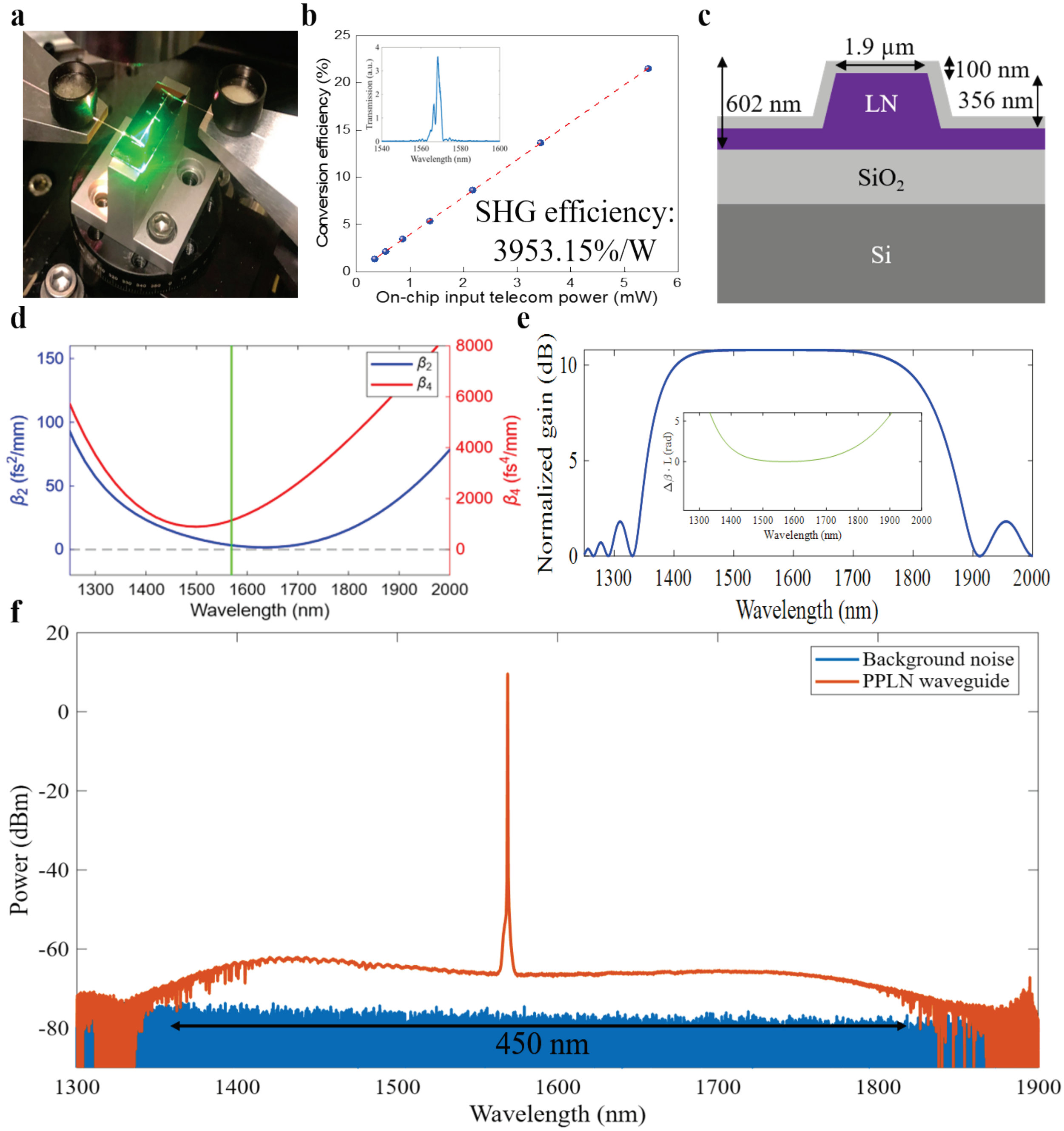

**Figure 2. On-chip broadband TFLN OPA measurements. (a)**, Picture of the OPA chip with the shining waveguide via high power NIR pumping. **(b)**, Linearly fitted SHG efficiency (3953.15%/W). Inset is the Measured SHG spectrum of the TFLN waveguide, enabled by the adaptive poling design. **(c)**, Cross-section view of TFLN waveguide. **(d)**, Simulated second-order and fourth-order dispersion coefficients $\beta_2$, and $\beta_4$ for our TFLN waveguide. By fitting the dispersion curves, we estimate the dispersion coefficients as $\beta_2 \approx 3.357$ fs$^2$/mm, and $\beta_4 \approx$ 1143.032 fs$^4$/mm, at fixed pump wavelength as the green line. **(e)**, Simulations of 3-dB Gain

bandwidth. We generate a broad and flat-top gain spectrum. The calculated OPA 3-dB gain bandwidth is 454 nm. Inset is the total phase mismatch simulation. Here the combination of near-zero second- and fourth-order dispersion $\beta_2$ and $\beta_4$ enables flat broadband gain profile with a fixed pump wavelength. **(f)**, Measured emission spectrum of the integrated OPA. A 3-dB bandwidth of the PPLN amplifier, reaching 450 nm, which is consistent with our simulation results in panel **(e)**. In addition, we also provide a summary comparison of 3-dB Gain bandwidth of integrated OPAs with literatures in Figure S1.

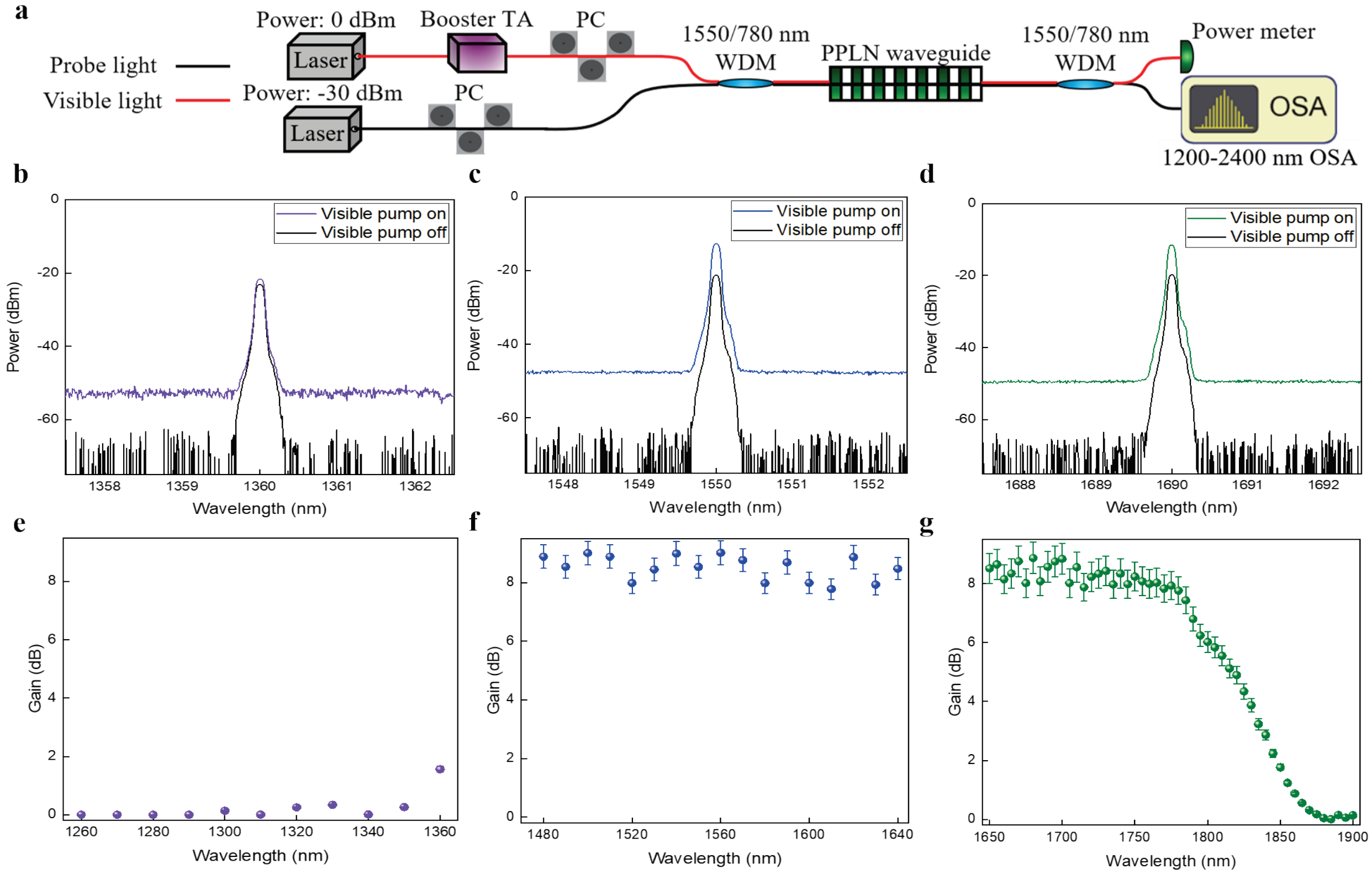

**Figure 3. Measurements of broadband amplification in PPLN waveguides via visible pumping. (a),** Experimental setup for CW-pumped OPA, where visible light is operated in high power regime to provide the gain for the probe light. We set the probe light to have -30 dBm power. The on-chip visible power is 74 mW, for a 24 dBm off-chip visible power. WDMs are used to combine and separate strong visible pump light and probe lights. **(b)** to **(d),** optical spectra of PPLN waveguide chip, with visible pump light on and off at 1360, 1550, and 1690 nm, respectively. The measured gains are 1.45 dB, 8.52 dB and 8.26 dB for these three wavelengths. Other selected optical spectra at different wavelengths are included in Figure S2. **(e)** to **(g),** gain measurements with probing light to be O-band, S-, C-, L-band, and in the 1650-1900 nm range, respectively. We collect more measurements for 1650-1900 nm range, as OPA gain measurements remain elusive so far for this wavelength range. The theoretical predictions of gain based on the SHG measurements in Figure 2(b) is 9.25 dB. Our measurements are consistent with theory and the emission spectrum in Figure 2(f), with maximum gain of 8.87 $\pm$ 0.39 dB. Wavelengths where the experimental gain is lower than the predicted gain are so because they fall outside the 3-dB amplification bandwidth of 450 nm. The theoretical calculation of the maximum gain is presented

in Figure S3. The error bars are determined by calculating the standard deviation from three measurements.

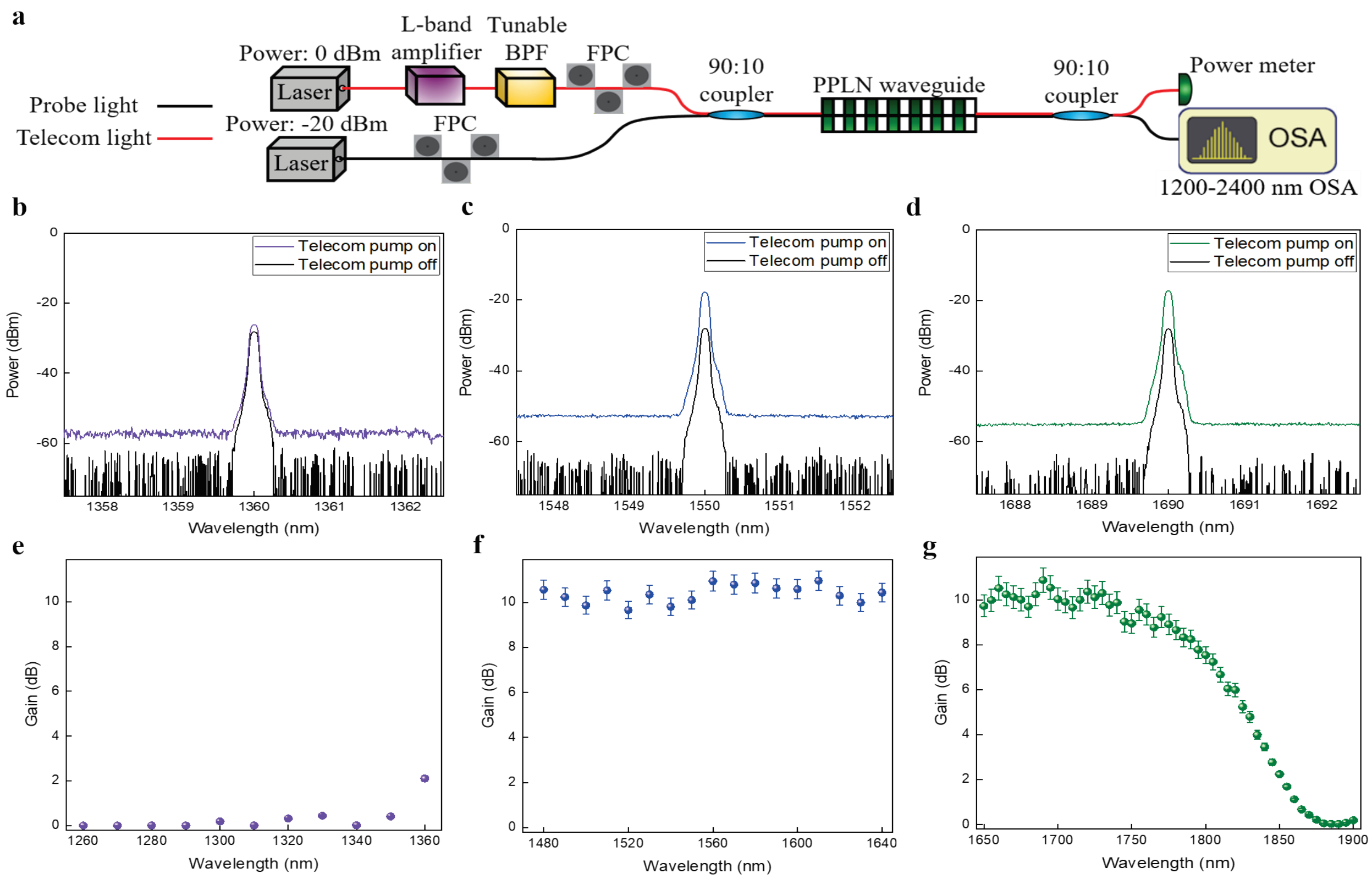

**Figure 4. Measurements of broadband amplification in PPLN waveguides via NIR pumping. (a)**, Experimental setup for the CW-pumped OPA, where telecom light is operated in the high power regime to provide the gain for the probing light. We set the probe light to have -20 dBm power. The on-chip telecom power is 170 mW, for a 26.5 dBm off-chip telecom power. 90:10 couplers are used to combine and separate telecom pump light and probe lights. **(b)** to **(d)**, optical spectra of the TFLN waveguide chip, with telecom pump light on and off at 1360, 1550, and 1690 nm, respectively. The measured gains are 2.05 dB, 10.17 dB and 10.81 dB for these three wavelengths. The other selected optical spectra at different wavelengths are included in Figure S5. **(e)** to **(g)**, gain measurements with probing light to be O-band, S-, C-, L-band, and in the 1650-1900 nm range, respectively. We collect more measurements for 1650-1900 nm range, as OPA gain measurements remain elusive so far for this wavelength range. The theoretical prediction of gain based on the SHG measurements in Figure 2(b) is 11.21 dB. To summarize, our measurements are consistent with theory and the emission spectrum in Figure 2(f), with maximum gain of 10.79 $\pm$ 0.43 dB, and the wavelengths where the experimental gain are lower than predicted gain is because they are outside the 3-dB amplification bandwidth of 450 nm. The theoretical calculation

of the maximum gain is presented in Figure S6. The error bars are determined by calculating the standard deviation from a set of three measurements.